\documentclass[5p,12pt,twocolumn,times]{elsarticle}
\usepackage{amsmath,amssymb,amsfonts}
\usepackage[T1]{fontenc}
\usepackage[utf8]{inputenc}
\usepackage{ulem}
\usepackage{xcolor}
\usepackage{lineno}
\usepackage{graphicx,float}
\usepackage[loose,bf,normalsize]{subfigure}
\usepackage[plainpages=false,pdfpagelabels,colorlinks=true,linkcolor=red,urlcolor=blue,citecolor=blue,pdfdisplaydoctitle=true,pdfduplex=DuplexFlipLongEdge]{hyperref}
\usepackage{hyperref}

\newcommand {\vk} {{\bf k}}
\newcommand \sa {\sigma}
\newcommand \sat {\tilde{\sigma}}

\newcommand \sot {\tilde{\bar{\sigma}}}
\newcommand \zt {\tilde{z}}
\newcommand \yt {\tilde{y}}
\newcommand \xt {\tilde{x}}

\newcommand \up {\uparrow}
\newcommand \pu {\downarrow}
\newcommand \epk[1] {\epsilon_{#1}}
\newcommand \epf {\epsilon^f}

\newcommand \ckm[1]  {c_{#1}^{\dag}}
\newcommand \ck[1]     {c_{#1}}
\newcommand \fim[1] {f_{#1}^{\dag}}
\newcommand \fis[1] {f_{#1}}

\newcommand \jhu {J_{H}}
\newcommand \med[1] {\langle{#1}\rangle}
\newcommand \vtr[1] {\mathbf{#1}}

\hypersetup{
pdfauthor = {C. Thomas, S. Burdin, and C. Lacroix},
pdftitle = {Metamagnetic transition due to Kondo effect in ferromagnetic systems},
}

\makeatletter
\def\ps@pprintTitle{%
 \let\@oddhead\@empty
 \let\@evenhead\@empty
 \def\@oddfoot{\rightline{\thepage}}%
 \let\@evenfoot\@oddfoot}
\makeatother

\begin{document}

\begin{frontmatter}

\title{Metamagnetic transition in the two $f$ orbitals Kondo lattice model}

\author[1]{Christopher Thomas}
\author[2]{S\'ebastien Burdin}
\author[3]{Claudine Lacroix}

\address[1]{Instituto de F\'isica, UFRGS, 91501-970 Porto Alegre, Brazil}
\address[2]{Universit\'e de Bordeaux, CNRS, LOMA, UMR 5798, F-33400 Talence, France}
\address[3]{Institut N\'eel, CNRS and Universit\'e Grenoble-Alpes, Boite Postale 166, 38042 Grenoble Cedex 09, France}

\begin{abstract}
 In this work, we study the effects of a transverse magnetic field in a Kondo lattice model with two $f$ orbitals interacting with the conduction electrons. The $f$ electrons that are present on the same site interact through Hund's coupling, while on neighboring sites they interact through intersite exchange. We consider here that part of  $f$ electrons are localized (orbital 1)  while another part (orbital 2) are delocalized, as it is frequent in uranium systems. Then, only electrons in the localized orbital 1 interact through exchange interaction with the neighboring ones, while electrons in orbital 2 are coupled with conduction electrons through a Kondo interaction. We obtain a solution where ferromagnetism and Kondo effect coexist for small values of an applied transverse magnetic field for $T\rightarrow0$. Increasing the transverse field, two situations can be obtained when Kondo coupling vanishes: first, a metamagnetic transition occurs just before or at the same time of the fully polarized state, and  second, a metamagnetic transition occurs when the spins are already pointing out along the magnetic field.
\end{abstract}

\end{frontmatter}

\section{Introduction}

The duality, local {\it versus} non-local character, of strongly correlated $f$-orbital electrons is a crucial microscopic quantum phenomenon. It generates unconventional macroscopic properties in rare earth and actinide heavy fermion compounds~\cite{Zwicknagl2002,Zwicknagl2003,Fulde2006}.
Among numerous fascinating examples, a family of uranium based compounds 
recently revealed the possibility of a new paradigm, where superconductivity and ferromagnetic order can coexist~\cite{Huxley2015,Aoki2019}. In these materials, application of a magnetic field perpendicular to the easy axis can generate a metamagnetic transition and a surprising enhancement of superconductivity inside a ferromagnetic phase. 

In order to understand the microscopic origin of this unanticipated behavior, we may first be inspired from the physics of heavy fermion cerium compounds, that can be described by a Kondo lattice model~\cite{Hewson1993}. In this case, the $4f^1$ Kondo ions can form a magnetically ordered lattice for small values of pressure where Ruderman-Kittel-Kasuya-Yosida (RKKY) interaction is predominant and go for a coherent Fermi Liquid state at higher pressure where local Kondo screening dominates. This pressure induced quantum transition can be well explained invoking Doniach's phase diagram~\cite{Doniach1977}. 
The itinerant contribution of {\it a priori} local $f$ electronic orbitals in the formation of the coherent Fermi liquid Kondo state is a signature of duality. It is also characterized by a large effective mass which can be revealed, for example, by big values of the specific heat Sommerfeld coefficient. On the other hand, $f$ electrons may be fully localized in the magnetically ordered state. 
Usually, both regimes are separated by a quantum critical point ~\cite{Si2010,Kirchner2020}, close to which a rich diversity of physical properties emerges, as for example, superconductivity in cerium based compounds~\cite{Zapf2001}. 

Revealing experimentally the passage from fully localized to itinerant and dual behavior is crucial and challenging. In CeRhIn$_5$, which is antiferromagnetic (AF) at ambient pressure, it was detected with de Haas-van Alphen experiment that these changes occur exactly at 2.4 GPa~\cite{Settai2001,Shishido2002}. It was observed that the large band signal indicates an increase of the effective mass from 5m$_0$ to 60m$_0$ which abruptly ``disappears'' as for a first order transition.
In CeRu$_2$Si$_2$, a clear Fermi surface reconstruction is observed along the metamagnetic transition~\cite{Aoki1993,Daou2006}, where a short Fermi surface takes places from the large Fermi surface, indicating a itinerant to localized behavior.
In YbRh$_2$Si$_2$~\cite{Guttler2021}, the authors studied the modification of the Fermi surface using high-resolution Compton scattering and showed a strong variation of the Fermi surface topology between low and high temperature regimes. They observed a clear enlargement of Fermi surface as temperature decreases, which is a signature of a coherent Kondo lattice ground state.  

On the other side, the actinides have a partially filled 5$f$ shell and their  behavior is different from the lanthanides behavior. At the beginning, the 5$f$ spatial wave function has a larger extent than the 4$f$. Furthermore, while the valence is 4$f^1$ in cerium based Kondo lattice systems, the  valence of uranium based compounds fluctuates between 5$f^2$ and 5$f^3$.  As a consequence, a microscopic phenomenological description of duality in these materials requires an adaptation of the Kondo lattice model, where magnetic ions are described by composite local $f^2$ multiplet states coupled to conduction electrons that account for the $f^2$-$f^3$ valence fluctuations~\cite{Perkins2007,Thomas2014}. Also, on general theoretical grounds, it is known that the possible presence of different channels for the conduction electrons might cause underscreeened or overscreened Kondo effect, depending on the comparison between the number of channels and the effective spin of the localized electronic multiplets~\cite{Nozieres1980}. In the uranium compounds, the $5f$ electrons are considered to be less localized than the $4f$ electrons in rare earth (lanthanides) compounds, allowing the possibility of underscreened Kondo effect, where the $5f$ electrons are only partially screened~\cite{Perkins2007,Thomas2014}. Some models consider the $5f^2$ pair of orbitals  as localized, while in other models one of the $5f$ electrons is itinerant,  which is necessary to describe the superconductivity in U-based compounds~\cite{Fidrysiak2019,Shick2019}, where $f$ electrons are considered to be the key for both superconductivity and ferromagnetism, when they exist. 

More precisly, in the group of the ferromagnetic superconductors, UGe$_2$~\cite{Saxena2000}, URhGe~\cite{Aoki2001}, and UCoGe~\cite{Huy2007}, superconductivity appears at low temperatures in the region where they are ferromagnetic. While for UGe$_2$ the superconducting state appears for high pressure, for URhGe and UCoGe the superconducting state already appears at ambient pressure and it persists for higher values of pressure. From the point of view of the magnetic response, the magnetic susceptibility $\chi_{a/b/c}$ is anisotropic for UGe$_2$, with $a-$axis being the easy magnetic axis, and for UCoGe, where $c-$axis is the easy magnetic axis. For URhGe, the evidence of anisotropy appears only below 50 K, where $\chi_b\neq \chi_c$ with $c$ being the easy axis. UGe$_2$ and UCoGe are considered to be Ising ferromagnets. For URhGe, the combination  between ferromagnetism along the c- and b-axes is the core of its extremely high field-reentrant superconductivity for magnetic fields applied along $b$.

The low value of magnetic moment per atom in UGe$_2$, URhGe, and UCoGe compounds (1.5, 0.4, and 0.06 $\mu_{B}/U$, respectively) indicates duality between localized and itinerant character of the $5f$ electrons in UGe$_2$~\cite{Troc2012}, while UCoGe has an itinerant behavior. This situation is also present in the uranium monochalcogenides US, USe and UTe~\cite{Schoenes1984}, although their Curie temperatures are much greater and they do not become superconductors at smaller temperature. 

A new fascinating compound, UTe$_2$, was recently added into the family of uranium superconductors~\cite{Ran2019}. However, differently from the previous one, UTe$_2$ does not order ferromagnetically. Nevertheless, the compound presents superconductivity coexisting with 
ferromagnetic fluctuations~\cite{Sundar2019}, and a rich phase diagram with different superconducting phases~\cite{Aoki2020}. What is also interesting is that the compound presents a metamagnetic transition for higher magnetic field, leading the system from paramagnetic to polarized paramagnetic~\cite{Miyake2019,Niu2020}. Earlier first principle calculations  
with local density approximation have shown  that UTe$_2$ was a Kondo semiconductor with small gap ($\sim 130$K) and flat bands around the Fermi Surface coming mainly from the 5$f$ electrons~\cite{Aoki2019a}. Other authors found a metallic ground state with moderate values of Coulomb repulsion ($\sim1.0$ eV) and an insulator-metal transition by increasing the Coulomb interaction~\cite{Ishizuka2019}. More recently~\cite{Shick2021}, 
density functional theory with exact diagonalization calculations indicate that $f-$valence in UTe$_2$ is close to 5$f^3$ (Coulomb interaction is about 3 eV).

The interplay between Kondo screening and ferromagnetism has been studied by different authors for spin-1/2 or spin-1 Kondo lattice models. Mean-field approaches result in phase diagrams where ferromagnetism is predominant for small interaction ($|J|/D$) and for the number of conduction electrons per site from zero to around 0.5, while Kondo phase appears for stronger interaction and in a larger range of conduction electrons density~\cite{Lacroix1979}. Other approaches report similar results~\cite{Fazekas1991}, without coexisting phases. The dynamical mean field theory calculations and further mean-field decouplings found coexisting phases~\cite{Li2010,Peters2012,Liu2013,Golez2013}, when  $f$-$f$ ferromagnetic exchange interaction is added to the Kondo lattice model~\cite{Thomas2011,Bernhard2015}. For URhGe, the density-matrix renormalization group calculations~\cite{Suzuki2019,Suzuki2020} also addressed this competition, but no coexistence is reported. In a recent paper~\cite{Bernhard2022} it was shown, using a fermionic mean field approximation, that under applied magnetic field a metamagnetic transition may occur in the Kondo lattice; in the present paper we study the effect of magnetic field on 2-$f$-orbitals Kondo lattice, which we think is more appropriate to U compounds which contain both localized and itinerant $f$-electrons.

One guideline for the theoretical work presented hereafter is URhGe and more precisely its magnetization, which displays a rapid variation when a transverse magnetic field is applied~\cite{Levy2005}.
In a more complete description, at low temperature, magnetic field perpendicular to the easy axis first destroys the superconducting phase around $2$ T, and at higher field a reentrant superconducting phase  appears between  $\sim 8$ T and $\sim 13$ T . Above the superconducting transition temperature, in the ferromagnetic phase  the spin-reorientation  occurs at $H_R\approx 12$ T. The maximum value of $T_{\mathrm{SC}}$ in the reentrant phase  coincides approximatively with $H_R$.
We study the effect of a transverse  magnetic field using a modified Kondo lattice model adapted for this uranium system.  Although a superconductiting phase is observed experimentally around the magnetization change, we do not explore this fact, and let it for a future work. This is justified since $T_{SC}$ is much smaller than the Curie temperature $T_{C}$ (respectively $0.25K$ and $9.5K$). 

The paper is organized as follows: In section \ref{Model} the model used to discuss the Kondo and ferromagnetic interactions in a transverse field is presented. The results and discussions are presented in sections \ref{Results} and \ref{Discussion}.

\section{The model}\label{Model}
We have in mind the objective of modeling uranium compounds with a valence fluctuating between 5$f^2$ and 5$f^3$. We will thus consider a Kondo lattice Hamiltonian where the Kondo ions are described by 5$f^2$ local multiplets. These two 5$f$ orbitals  (labeled as orbitals $1$ and $2$) are centered on each lattice site $i$ and  we define the spin of each of these electrons as $\vtr{S}_{i1}$ and $\vtr{S}_{i2}$. The intrasite interaction should in principle include both the local Coulomb repulsion and the Hund's coupling. However, by enforcing the localization and describing the corresponding degrees of freedom with effective spin operators, the Coulomb repulsion is implicitly assumed to be strong. 

\begin{figure}[H]
  \centering
  \includegraphics[width=1\columnwidth]{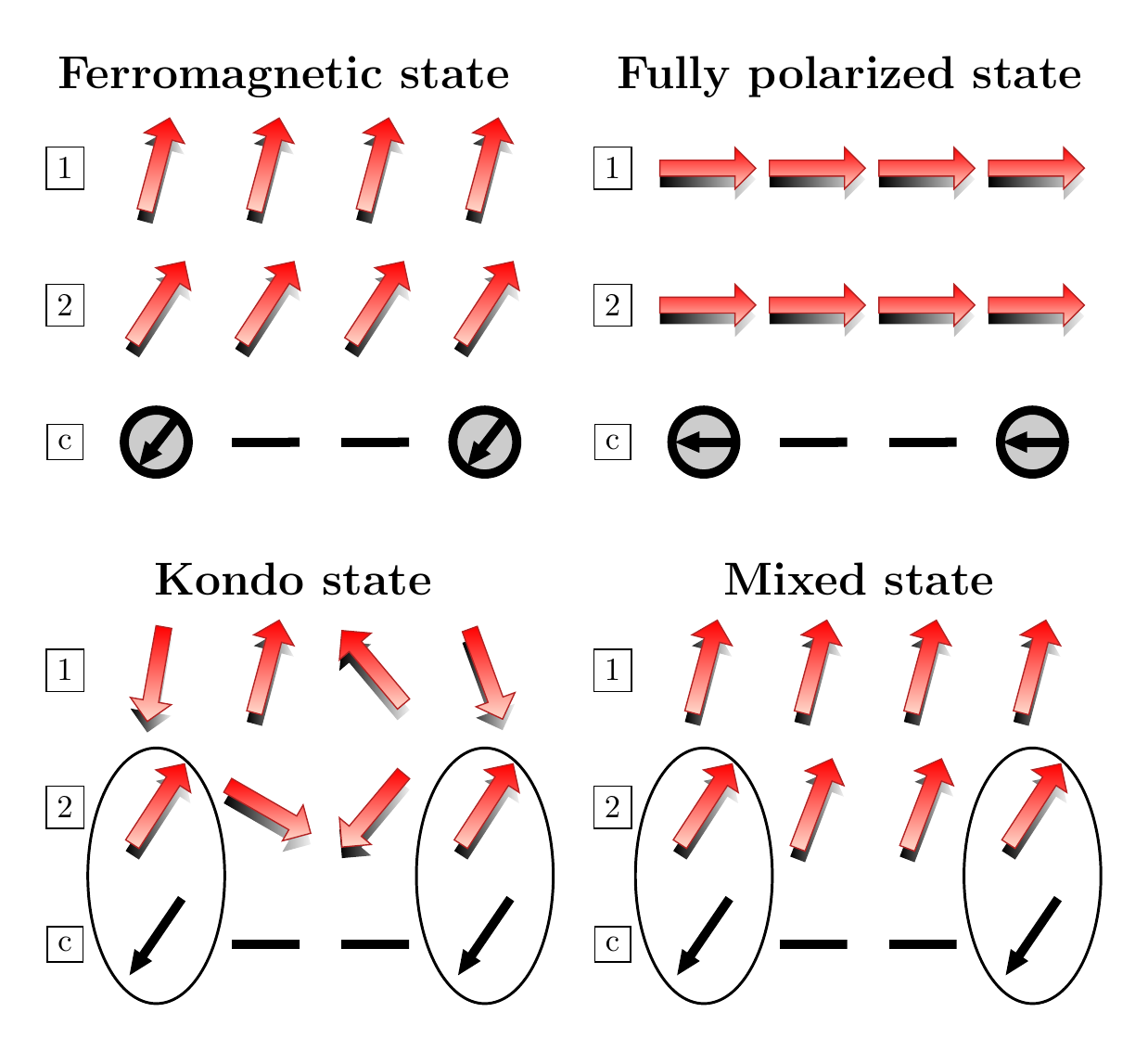}
  \caption{\label{fig:sche_phases} Some possible ordered phases of the system. The moments of the 3 types of electrons ( $f_{1}$,  $f_{2}$ and $c$ electrons)  are indicated separately.  From left to right, from top to bottom. Ferromagnetic phase: the averaged magnetizations for $f-$ and $-c$ electrons is non zero. Fully Polarized phase : particular ferromagnetic phase where the magnetizations are saturated and aligned along the external magnetic field; the $f-$electron magnetisation per site is $S=1$. The $c$ electrons couple antiparallel to the $f$ ones since local Kondo exchange is negative. Pure Kondo state: $f-$electrons in orbital 2 and $c-$electrons form  Kondo singlets, and the $f-$electrons in orbital 1 are paramagnetic. Mixed state: coexistence of ferromagnetic order (non-zero magnetization) and Kondo screening of electrons in orbital 2.} 
\end{figure}

Furthermore, we also want to describe the duality of the localized $f$-electrons. 
We consider here that part of $f$-electrons is fully localized (orbital $1$) while another part (orbital $2$) can be partially delocalized, as it is frequent in U systems. Then, only electrons in the fully localized orbital $1$ are supposed to interact through  ferromagnetic intersite exchange interaction, while electrons in orbital $2$ are coupled with conduction electrons through a Kondo interaction.  
The itinerant electrons emerge from the $f^2$-$f^3$ valence fluctuation effects as well as from other  electronic bands.
In figure \ref{fig:sche_phases} we can see some possible solutions that can be described by the proposed model. In the ferromagnetic phase  spins with all components ($c$, and both $f$ orbitals) exhibit a non-zero magnetization. In the fully polarized phase, all spins are aligned  with the transverse magnetic field. In the Kondo phase, the spin of the $f$-electrons in orbital 2 couple with the spin of the conduction electrons through Kondo interaction and the spins of orbital 1 remain paramagnetic and finally, the mixed state where Kondo and ferromagnetism are both present.

The generalized Kondo lattice Hamiltonian of the system is written as: 

\begin{align}
H&=H_{\mathrm{intersite}}+H_{\mathrm{local}}+H_{\mathrm{Kondo}}+H_c~, 
\label{Hamiltonian}
  \end{align}
where the first term is intersite exchange that we consider here as an Ising interaction between localized spins in orbital 1: 
\begin{align}
H_{\mathrm{intersite}}&=-J\sum_{\med{ij}}S_{i1}^zS_{j1}^z\,~. 
\end{align}
This Ising exchange interaction describes  phenomenologically the large magnetic uniaxial  anisotropy observed experimentally in the uranium based ferromagnetic superconductors. 
The second term in the Hamiltonian includes the on-site interactions: Hund's coupling ($J_H\ge 0$) between electrons in orbitals $1$ and $2$, and applied magnetic field, which can be in any direction:
 
 \begin{align}
 H_{\mathrm{local}}&=-J_{\mathrm{H}}\sum_i\vtr{S}_{i1}\cdot\vtr{S}_{i2}-\vtr{h}\cdot\sum_i \big(\vtr{S}_{i1} + \vtr{S}_{i2}\big)\,~.
\end{align}

The third term is the local Kondo coupling ($J_K\ge 0$) between $f-$electrons in orbital 2 and $c-$electrons:  
\begin{align}
  H_{\mathrm Kondo}&=J_K\sum_i\vtr{S}_{i2}\cdot\vtr{s}_i\,~, 
\end{align}
where $\vtr{s}_i$ denotes the local spin density of the $c-$electrons. 
The last term describes the conduction band: 
\begin{align}
   H_c&=\sum_{\vk\sa}(\epk{\vk}-\mu)\ckm{\vk\sa}\ck{\vk\sa}+N\mu n_c\,~, 
\end{align}
where $\sa=\uparrow, \downarrow$ is the spin component and $\vk$ is the electron momentum. $\epk{\vk}$ is the dispersion relation, $\mu$ is the chemical potential associated with the averaged total number of conduction electrons per site $n_c$, and $N$ denotes the number of lattice sites. 

In the following, we will consider that there are 2 $f$ electrons on each site, one in each orbital. Thus in the absence of Kondo interaction, these two electrons are coupled  by Hund's interaction in a $S=1$ state  and they interact with neighboring sites through Ising exchange. The problem is solved in two steps: first, we solve the $f-$electrons problem, with a magnetic field applied in the $x-z$ plane; second, the Kondo coupling  is added for orbital 2.

\subsection{Solution for the magnetic part}

We  first treat the effect of the Ising and Hund's interactions in orbitals 1 and 2 in the presence of magnetic field. 
\begin{align}\label{eq:h1m}
H_f\equiv H_{\mathrm{intersite}}+H_{\mathrm{local}}\,,
\end{align}
where the different terms are defined above. 

In a mean field decoupling , we introduce the effective field acting on electrons in orbital 1 and 2 respectively, $h_{w1}$ and $h_{w2}$. The purely $f$-orbital part of the Hamiltonian is thus approximated as: 

\begin{align}
H_f\approx   H^{\mathrm{MF}}_f
\equiv E_0-\sum_i\left( 
\vtr{h}_{w1}\cdot\vtr{S}_{i1}+\vtr{h}_{w2} \cdot\vtr{S}_{i2}\right)\,,\label{eq:hmf1b}
\end{align}
where  $E_0=N\frac{J'}{2}(m_1^{z})^2+N\jhu(m_1^{x}m_2^{x}+m_1^{z}m_2^{z})$ 
and $m_{1(2)}^{z}=\frac{1}{N}\sum_{i}\med{S_{i1(2)}^{z}}$ and $m_{1(2)}^{x}=\frac{1}{N}\sum_{i}\med{S_{i1(2)}^{x}}$ are the components of magnetization  in the $z$ and $x$ directions.  Note the redefinition $J'=2zJ$, where $z$ is the number of nearest neighbors. 

The  effective fields are defined as
\begin{align}\label{eq:hw11}
\vtr{h}_{w1}=\left(
{\begin{array}{c}
h_x+\jhu m_2^{x} \\ 
0 \\ 
h_z+J'm_1^{z}+\jhu m_2^{z}
\end{array}}
\right)\,,
\end{align}

\begin{align}\label{eq:hw22}
\vtr{h}_{w2}=\left(
{\begin{array}{c}
h_x+\jhu m_1^{x} \\ 
0 \\ 
h_z+\jhu m_1^{z}
\end{array}}
\right)\,.
\end{align}
We can also identify the angles of  magnetization for both orbitals 1 and  2 with respect to x-direction:
\begin{align}
\tan{\theta_1}&=\frac{m_1^{z}}{m_1^{x}}=\frac{h_z+J'm_1^{z}+\jhu m_2^{z}}{h_x+\jhu m_2^{x}}\,,\label{eq:th1}\\
\tan{\theta_2}&=\frac{m_2^{z}}{m_2^{x}}=\frac{h_z+\jhu m_1^{z}}{h_x+\jhu m_1^{x}}\,.\label{eq:th2}
\end{align}
These definitions help us to have a more compact form for the set of self-consistent mean-field equations, that can be written as
\begin{align}
m_1^{z}&=\frac{\sin{\theta_1}}{2}\tanh{\frac{\beta\|\vtr{h}_{w1}\|}{2}}\,,\label{eq:m1fz}\\
m_1^{x}&=\frac{\cos{\theta_1}}{2}\tanh{\frac{\beta\|\vtr{h}_{w1}\|}{2}}\,,\\
m_2^{z}&=\frac{\sin{\theta_2}}{2}\tanh{\frac{\beta\|\vtr{h}_{w2}\|}{2}}\,,\\
m_2^{x}&=\frac{\cos{\theta_2}}{2}\tanh{\frac{\beta\|\vtr{h}_{w2}\|}{2}}\,.\label{eq:m2fx}
\end{align}

The problem can be solved considering the external magnetic field in the $x$ - $z$ plane. However, as we will consider the  Kondo coupling between the $f$ electrons in orbital 2 with the conduction electrons, the magnetization of orbital 2, $m_2^f$, has to be calculated self-consistently  in the presence of Kondo coupling. This is  presented below. 

\subsection{Solution in the presence of Kondo coupling}
We now consider the full Hamiltonian defined by eq.~(\ref{Hamiltonian}), where the purely $f$-orbital part is approximated by the mean-field expression eq.~(\ref{eq:hmf1b}): 
\begin{align}
H\approx H^{\mathrm{MF1}}\equiv E_0+H^{\mathrm{MF}}_f + H_{\mathrm{Kondo}}+H_c~, 
\end{align}
First it is necessary to make a basis change of the spin quantization axis: in the presence of the external magnetic field, the $f$ magnetic moments tend to point  towards the direction of magnetic field, but they are not aligned with field since the Ising exchange acts as a local anisotropy (see above eqs. \eqref{eq:m1fz}-\eqref{eq:m2fx}). For this reason, we will change the quantization axis for the spins.  We use the symbol tilde to represent this new quantization axis:  $\tilde{z}$   and $\tilde{\sigma}$ correspond to the $z$ axis rotated by angle ${\theta_2}$ defined in the previous section.
Invoking the spin rotational invariance of $H_{\mathrm{Kondo}}$ and $H_c$, the mean-field Hamiltonian is thus rewritten as 
\begin{align}
H^{\mathrm{MF1}}
=& E_0+N\mu n_c -\vtr{h}_{w1}\cdot\sum_i\vtr{S}_{i1}
-h_{w2}\sum_i S_{i2}^{\zt}\nonumber\\
&+
J_K\sum_i\tilde{\vtr{S}}_{i2}\cdot\tilde{\vtr{s}}_i
+
\sum_{\vk\tilde{\sa}}(\epk{\vk}-\mu)\ckm{\vk\tilde{\sa}}\ck{\vk\tilde{\sa}}~. \label{eq:h2til}
\end{align}

The Kondo effect is treated within usual mean field approximation using fermionic representation of the spin operators : 
\begin{align}
S_{i2}^{\zt}=&\frac{1}{2}\left( 
f_{i2\tilde{\up}}^\dagger f_{i2\tilde{\up}}-f_{i2\tilde{\pu}}^\dagger f_{i2\tilde{\pu}}\right)\\
S_{i2}^{\yt}=&\frac{i}{2}\left( 
f_{i2\tilde{\pu}}^\dagger f_{i2\tilde{\up}}-f_{i2\tilde{\up}}^\dagger f_{i2\tilde{\pu}}\right)\\
S_{i2}^{\xt}=&\frac{1}{2}\left( 
f_{i2\tilde{\pu}}^\dagger f_{i2\tilde{\up}}+f_{i2\tilde{\up}}^\dagger f_{i2\tilde{\pu}}\right)~, 
\end{align}
where the Abrikosov fermion annihilation (creation) operators $f_{i2\tilde{\sigma}}^{(\dagger)}$ satisfy the local constraints $\sum_{\tilde{\sigma}}f_{i2\tilde{\sigma}}^{\dagger}f_{i2\tilde{\sigma}}=1$. Within the mean-field approximation, this constraint is satisfied on average,  $\sum_{i\tilde{\sigma}}\med{f_{i2\tilde{\sigma}}^{\dagger}f_{i2\tilde{\sigma}}}=N$, by introducing an effective energy level $\epf_2$. Finally the mean-field approximations for the Hamiltonian give \\
\begin{align}\label{eq:hmf2a}
H\approx H^{\mathrm{MF2}}&=
E_0^{\mathrm{MF2}} -\vtr{h}_{w1}\cdot\sum_i\vtr{S}_{i1}
+\sum_{i\sat}\epf_{2\sat}\fim{i2\sat}\fis{i2\sat}
\nonumber\\
&+\sum_{i\sat} \Lambda_{2}^{\sot}\Big(\ckm{i\sat}\fis{i2\sat}
+\fim{i2\sat}\ck{i\sat}\Big)+\sum_{\vk\sat}\epk{\vk}^c\ckm{\vk\sat}\ck{\vk\sat}\,,
\end{align}
where $\lambda_{i2}^{\sat}=\med{\fim{i2\sat}\ck{i\sat}}$ and
\begin{align}
E_0^{\mathrm{MF2}}&= E_0+N\mu n_c-N\epf_2+J_KN\lambda_{2}^{\tilde{\up}}\lambda_{2}^{\tilde{\pu}}\,,\\
\epf_{2\sat}&=\epf_2-\sat h_{w2}\,,\\
\Lambda_{2}^{\sat}&=-\frac{J_K}{2}\lambda_{2}^{\sat}\,,\\ 
\epk{\vk}^c&=\epk{\vk}-\mu\label{eq:relationLambdalambda}\,.\\
\end{align}
$\sat=\pm 1/2$ and $h_{w2}=\|\vtr{h}_{w2}\|$.
With the first mean-field approximation that we used, the momenta $\vtr{S}_{i1}$ describing the local electronic orbital 1 
are effectively decoupled from both the orbital 2 and the conduction electrons. However, the Hund interaction between orbitals 1 and 2 is still implicitly present through the effective fields $\vtr{h}_{w1}$ and $\vtr{h}_{w2}$ given by 
solving eqs.~(\ref{eq:hw11}) and~(\ref{eq:hw22}) and the self-consistent relations~(\ref{eq:m1fz}) and~(\ref{eq:m2fx}). 
The second mean-field approximation replaces the Kondo interaction by an effective hybridization between  orbital 2 and  conduction electrons. It describes qualitatively and quantitatively the Kondo-singlet correlations that can occur at low temperature. 
The mean-field Hamiltonian can be diagonalized using the space momentum representation $\vk$: 
{\small
\begin{align}\label{eq:hmf2b}
H^{\mathrm{MF2}}&=
E_0^{\mathrm{MF2}} -\vtr{h}_{w1}\cdot\sum_i\vtr{S}_{i1}
+
\sum_{\vk\sat}\left( \ckm{\vk\sat}, \fim{\vk 2\sat}\right) H_{\vk\sat}^{cf_2}\left(
\begin{array}{c}
\ck{\vk\sat}\\
\fis{\vk 2\sat}
\end{array}
\right)~, 
\end{align}}
with 
\begin{align}
H_{\vk\sat}^{cf_2}\equiv
\left(
\begin{array}{cc}
\epk{\vk}^c & \Lambda_{2}^{\sat}\\
~~&~~\\
\Lambda_{2}^{\sat} & \epf_{2\sat}
\end{array}
\right)~. 
\end{align}

From $H^{\mathrm{MF2}}$ and using the above expression of the block $H_{\vk\sat}^{cf_2}$, 
the one body Green's functions for orbital 2 and  conduction electrons can be written as
\begin{align}
g_{\vk\sat}^{cc}(i\omega)&=\frac{i\omega-\epf_{2\sat}}{(i\omega-\epk{\vk}^c)(i\omega-\epf_{2\sat})-(\Lambda_2^{\sot})^2} 
\,,\\
g_{\vk\sat}^{f_2f_2}(i\omega)&=\frac{i\omega-\epk{\vk}^c}{(i\omega-\epk{\vk}^c)(i\omega-\epf_{2\sat})-(\Lambda_2^{\sot})^2}
\,,\\
g_{\vk\sat}^{cf_2}(i\omega)&=g_{\vk\sat}^{f_2c}(i\omega)=\frac{-\Lambda_2^{\sot}}{(i\omega-\epk{\vk}^c)(i\omega_n-\epf_{2\sat})-(\Lambda_2^{\sot})^2}
\,. 
\end{align}

From the Green functions we can calculate the self-consistent parameters: 
\begin{align}\label{eq:ncf_lambda}
n_{\sat}^c&=\sum_{\vk}\int_{-\infty}^{\infty}d\omega f(\omega) \rho_{\vk\sat}^c(\omega)\,,\\
\label{eq:ncf_lambdabis}n_{\sat}^{f_2}&=\sum_{\vk}\int_{-\infty}^{\infty}d\omega f(\omega) \rho_{\vk\sat}^{f_2}(\omega)\,,\\
\label{eq:ncf_lambdater}\lambda_2^{\sat}&=\sum_{\vk}\int_{-\infty}^{\infty}d\omega f(\omega) \rho_{\vk\sat}^{cf_2}(\omega)\,.
\end{align}
where the the spectral density functions are
\begin{align}
\rho_{\vk\sat}^c(\omega)&=-\frac{1}{\pi} \Im{\big[g_{\vk\sat}^{cc}(i\omega)\big]}\,,\\
\rho_{\vk\sat}^{f_2}(\omega)&=-\frac{1}{\pi} \Im{\big[g_{\vk\sat}^{f_2f_2}(i\omega)\big]}\,,\\
\rho_{\vk\sat}^{cf_2}(\omega)&=-\frac{1}{\pi} \Im{\big[g_{\vk\sat}^{cf_2}(i\omega)\big]}\,.
\end{align}

The parameters defined in equations~\eqref{eq:ncf_lambda}\eqref{eq:ncf_lambdabis}\eqref{eq:ncf_lambdater} are solved considering a square band (i.e., a constant density 
of states) for the conduction electrons. The only $\vk$ dependence comes from the dispersion relation $\epk{\vk}$, in this way, the sum over $\vk$ is changed by an integral over energy as $\sum_{\vk}\rightarrow \int_{-D}^{D}\rho_0(\epsilon) d\epsilon$. The non-interacting conduction electrons density of states is taken as $\rho_0=1/2D$ in the interval $[-D:D]$. $D$ is the half of the value of the bandwidth.

Using the above expressions  for the Green's functions and invoking the relation eq.~(\ref{eq:relationLambdalambda}), 
the mean-field self-consistent equations can be rewritten as 
{\footnotesize
\begin{align}
  n_{\sat}^c&=\rho_0\int_{-D}^D
d\epsilon
  \frac{\left( E_+^{\sat}(\epsilon)-\epf_{2\sat}\right) f(E_+^{\sat}(\epsilon))
   -\left( E_-^{\sat}(\epsilon)-\epf_{2\sat}\right) f(E_-^{\sat}(\epsilon))
  }{\Delta E(\epsilon)}
 \,,\label{eq:ncsa}
\end{align}}
{\footnotesize
\begin{align}
n_{\sat}^{f_2}&=\rho_0\int_{-D}^D
d\epsilon
\frac{\left( E_+^{\sat}(\epsilon)-\epk{\vk}^c\right) f(E_+^{\sat}(\epsilon))
- \left( E_-^{\sat}(\epsilon)-\epk{\vk}^c\right) f(E_-^{\sat}(\epsilon))}{\Delta E(\epsilon)}
\,,\label{eq:nf2tsa}
\end{align}}
\begin{align}
\Lambda_2^{\sat}&=\frac{\rho_0 J_K}{2}
\Lambda_2^{\sat}
\int_{-D}^D
d\epsilon
\frac{f(E_+^{\sat}(\epsilon))- f(E_-^{\sat}(\epsilon))}{\Delta E(\epsilon)}\,.\label{eq:l2sasa}
\end{align}
where $f(\omega)=(1+e^{\beta\omega})^{-1}$ denotes the Fermi-Dirac function, and 
\begin{align}
E_{\pm}^{\sat}(\epsilon)&\equiv\frac{\epsilon+\epf_{2\sat}\pm\sqrt{(\epsilon-\epf_{2\sat})^2+4(\Lambda_2^{\sot})^2}}{2}\,,\\
\Delta E(\epsilon)&\equiv E_2^{\sat}(\epsilon)-E_3^{\sat}(\epsilon)~. 
\end{align}
We note that the mean-field equation~(\ref{eq:l2sasa}) has a trivial solution $\Lambda_2^{\sat}=0$ which is realized in the non-Kondo phases where the local $f$ and the conduction electrons are decoupled. When a solution $\Lambda_2^{\sat}\neq 0$ exists and is energetically stable, a Kondo phase is realized. On top of this usual mean-field description of Kondo effect, we also consider here the possibility of magnetic ordering, which may coexist or not with the Kondo effect. 
From equation \eqref{eq:nf2tsa} we can obtain the magnetization of orbital 2 in the rotated direction (noted with tilda), 
$M_{2}^{f\zt}=\frac{1}{2}(n_{\tilde{\up}}^{f_2}-n_{\tilde{\pu}}^{f_2})$. However, this magnetization is in the direction of the effective field $h_{w2}$ and not in the original $z$ or $x$ directions. 
The magnetization in the initial cartesian coordinates is thus given by 
\begin{align}
m_2^{z}&=M_{2}^{f\zt}\sin{\theta_2}\,,\label{eq:m2fz}\\
m_2^{x}&=M_{2}^{f\zt}\cos{\theta_2}\,,
\end{align} 
where $\theta_2$ was defined previously in equation \eqref{eq:th2}.

Finally, we can solve self-consistently our set of six equations \eqref{eq:ncsa}, \eqref{eq:nf2tsa}, \eqref{eq:m1fz}, \eqref{eq:m2fz}, and \eqref{eq:l2sasa}, and determine the parameters $\mu$, $\epf_2$, $m_1^{z}$, $m_2^{z}$, $\tilde{\lambda}_2^{\up}$, and $\tilde{\lambda}_2^{\pu}$, respectively. We have checked  the self-consistent solutions looking for the parameters that minimize the mean-field Hamiltonian presented in equation \eqref{eq:hmf2b} for the case $T\rightarrow0$.
The numerical results are presented in the next section.

\section{Results}\label{Results}

We explore the phase diagram for various sets of the model's parameters. First, we fix  the number of particles in orbital 2, $\med{n_{2f}}\equiv1$, and the number of conduction electrons $\med{n_c}\equiv n_c = 0.8$, with the help of the auxiliary Lagrangian's multipliers, $\epf_2$ and $\mu$. In the last part, we will explore the effect of variation of $n_c$. The magnetic field is taken in the $x$ direction, fixing $h_z=0$. The energies are scaled with respect to the half bandwidth $D=1$, using $k_{\mathrm{B}}=1$ and $T=0.0001$ for all figures. 

\subsection{\label{Section:result-field-effect}Effect of  magnetic field and \texorpdfstring{$J_K$ for $J'=J_H=0$}{JK for J'=JH=0}.}\label{sec:j0jh0}
First we study here the effect of a magnetic field and Kondo coupling in the absence of both Hund and intersite interactions, $J_H=0$ and $J'=0$. 
In this case the local spins $\vtr{S}_{i1}$ are fully decoupled from the orbital 2 and the conduction electrons. The model corresponds to an usual Kondo lattice for orbital 2, and we use the Abrikosov fermions $f_{i2}$ to describe the local spins $\vtr{S}_{i2}$. Since rotational symmetry is preserved at zero field when $J'=0$, we arbitrarily choose the $x$-axis along the direction of the applied magnetic field. 

\begin{figure}[H]
  \centering
  \includegraphics[width=1\columnwidth]{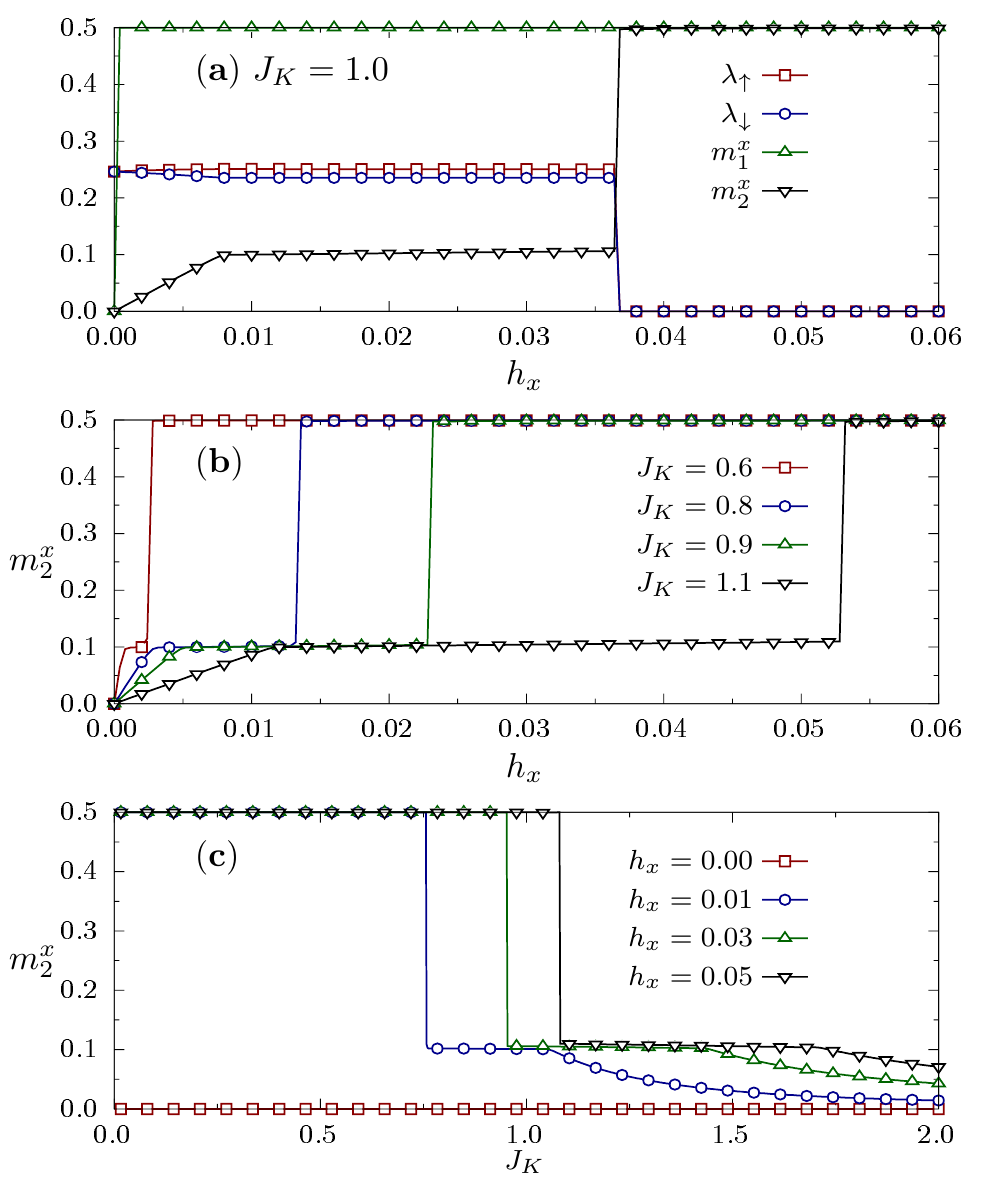}
  \caption{\label{fig:lambda_m2x_jk} ($\mathbf a$) Variation of the magnetization and Kondo mean field parameter as a function of $h_x$. $\lambda_{\up}\neq\lambda_{\pu}$  while $m_2^x$ is finite and not saturated. ($\mathbf b$)  The magnetization $m_2^x$ as a function of $h_x$. Higher values of $J_K$ allows longer plateau behaviour. ($\mathbf c$) $m_2^x$ as a function of $J_K$. At the same point where $\lambda_{\up}$ and $\lambda_{\pu}$ have a discontinuity, $m_2^x$ has a step from its saturated value to  $m_2^x=(1-n_c)/2$ indicating the mixed phase. It goes continually to zero for large values of $J_K$ . $m_1^z$ and $m_2^z$ are zero for all figures since $J'=J_{H}=0$. } 
 \end{figure}

The figure \ref{fig:lambda_m2x_jk}$(\mathbf{a})$ shows the variation of the effective Kondo hybridizations $\lambda_{\sigma}$ and the magnetizations for both orbitals as a function of the magnetic field $h_x$, for $J_K=1.0$ and $J=J_H=0$. Here, orbital 1 is fully decoupled from the other electrons, therefore in the ground state its magnetization saturates to its maximal value as soon as $h_x\neq0$. The magnetization of orbital 2 is more interesting and we can identify three regimes depending on intensity of the applied magnetic field: 
at relatively small $h_x$, we find that $m_2^x$  increases linearly with $h_x$, revealing a Fermi-liquid regime with a finite susceptibility that correspond to the usual Kondo coherent regime. On the other side, the magnetization $m_2^x$ saturates to its maximal value when the applied field is higher than a critical value ${h_K^{\star}}$. ${h_K^{\star}}$ is defined as the critical field necessary to destroy the Kondo hybridization and it coincides with the saturation of $m_2^x$ only in some cases (see section \ref{subsec:hund}).  The intermediate regime of magnetic fields, above the linear response and below the critical value ${h_K^{\star}}$, we find a magnetization plateau at the value $m_2^x\approx (1-n_c)/2$ independently from $h_x$ and $J_K$ as can be seen in figures \ref{fig:lambda_m2x_jk}({\bf b}) and \ref{fig:lambda_m2x_jk}({\bf c}).

The three magnetization regimes described above can be interpreted as follows: for relatively small $h_x$, the local Kondo spins $\vtr{S}_{i2}$ and the conduction electrons form a coherent Kondo Fermi liquid state characterized by a constant susceptibility and a linear magnetization. The occurrence of an intermediate plateau regime at higher values of the field was predicted previously for a Kondo lattice (see Figure 5 in~\citep{Burdin2009}). It can be explained using a strong Kondo coupling picture: independently from the magnetic field, in  this Kondo phase a fraction $n_c$ of the Kondo spins are screened and form local Kondo singlets with the conduction electrons. The remaining fraction $1-n_c$ of unscreened Kondo spins can thus be magnetized without breaking the macroscopic coherent Kondo state. This intermediate Kondo regime with a magnetization plateau is energetically favorable as long as the Kondo singlet energy (typically the Kondo temperature $T_K$) is higher than the Zeeman energy (proportional to $h_x$).

This plateau corresponds to a Kondo phase with $\lambda_{\up}$ and $\lambda_{\pu}$ slightly different from each other but both non-zero~(see figure \ref{fig:lambda_m2x_jk}(${\mathbf a}$)). In the Kondo phase, $\lambda_{\up}$ and $\lambda_{\pu}$ take relatively close values that differ as soon as $h_x$ is non-zero and both vanish above the same critical field ${h_K^{\star}}$ which marks the breakdown of Kondo effect. 

\begin{figure}[H]     
    \centering
    \includegraphics[width=1\columnwidth]{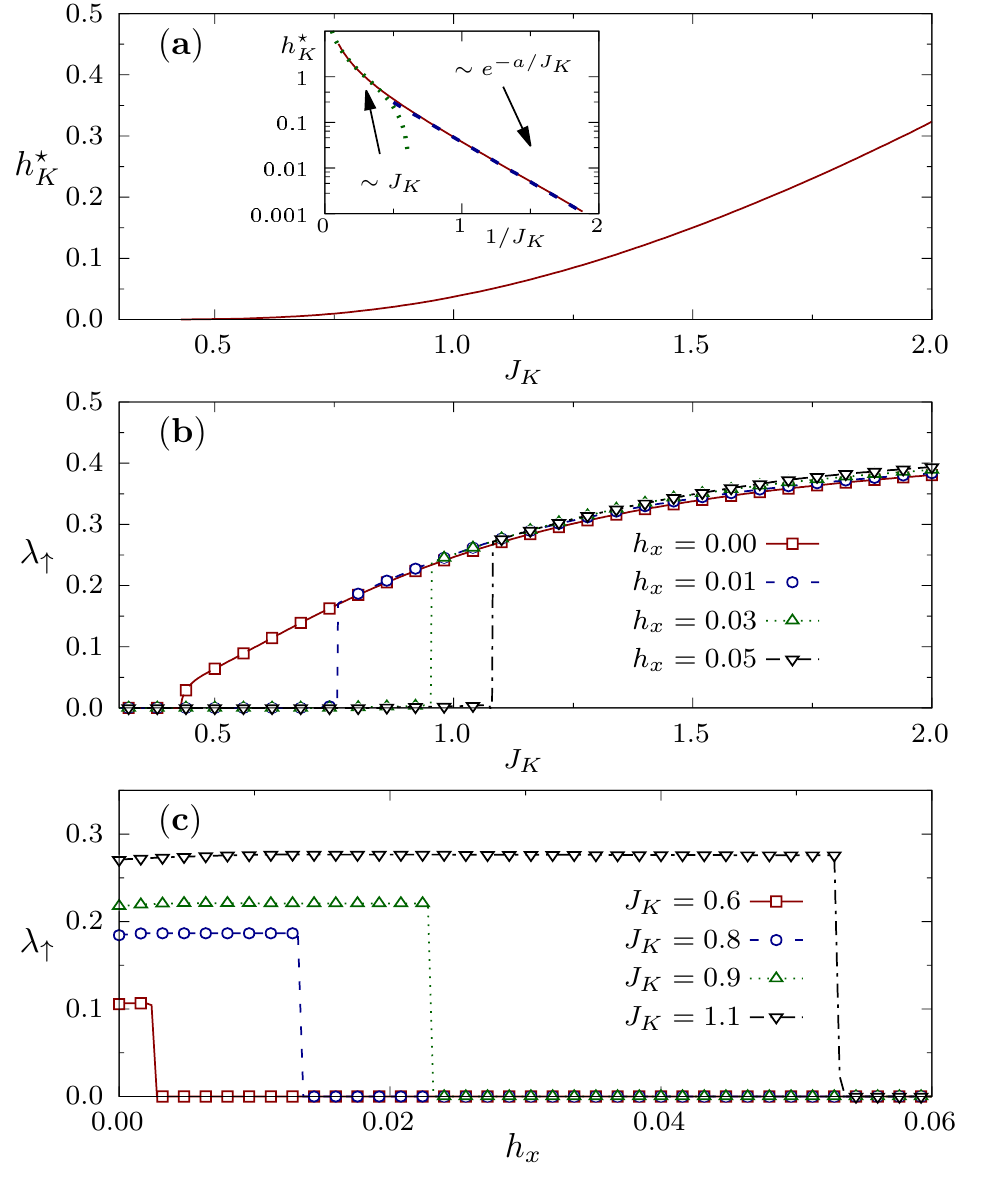}
    \caption{\label{fig:hxc_lambda} ({\bf a}) The critical magnetic field, $h_K^{\star}$, as a function of $J_K$. The critical field follows an exponential behavior ($\sim\exp(-a/J_K)$) for small values of $J_K$ (inset). For higher values, the $J_K$ dependence is linear, as it can also be seen in the inset. ({\bf b}) $\lambda_{\uparrow}$ as a function of $J_K$ for different values of $h_x$. The effective hybridization is zero for small values of $J_K$. For $h_{x}\neq0$, there is a critical value of $J_K$ where $\lambda_{\up}$ increases abruptly. 
      ({\bf c}) $\lambda_{\up}$ as a function of $h_x$ for different values of $J_K$. As $J_K$ increases, the absolute value of $\lambda_{\up}$ increases and the field $h_x$ above which Kondo pairing is destroyed increases also. $J'=J_{H}=0$ for all figures.}
  \end{figure}

The variation of ${h_K^{\star}}$ is depicted in figure~\ref{fig:hxc_lambda}({\bf a}) and in its inset. After an exponential behavior ($\sim\exp(-a/J_K)$) at small Kondo coupling, it goes linearly 
with $J_K$ for strong Kondo interaction. From the inset, it is possible to verify that the Kondo critical field scales as $T_K$ for $h_x=0$ \citep{Burdin2009}. The figures~\ref{fig:hxc_lambda}({\bf b}) and~\ref{fig:hxc_lambda}({\bf c}) present the variation of the $f-c$ effective hybridization $\lambda_{\up}$ as a function of $J_K$ for fixed values of $h_x$, and as a function of $h_x$ for fixed values of $J_K$, respectively. 
We find that $\lambda_{\up}$ vanishes at small $J_K$ as soon as $h_x$ is non-zero, it abruptly jumps to a finite value around a critical value of $J_K$ and then increases continuously with $J_K$. 
When fixing the Kondo coupling, we find that the effective hybridization  is almost constant as $h_x$ increases and vanishes abruptly at $h_x\equiv h_K^{\star}$. This is consistent with the abrupt jump observed when increasing $J_K$, reflecting the fact that $h_K^{\star}$ depends on $J_K$ in a monotonous way.

\subsection{Effect of the intersite magnetic interaction \texorpdfstring{$J'$ for $J_H=0$}{J' for JH=0}}
\begin{figure}[H]
    \centering
\includegraphics[width=1\columnwidth]{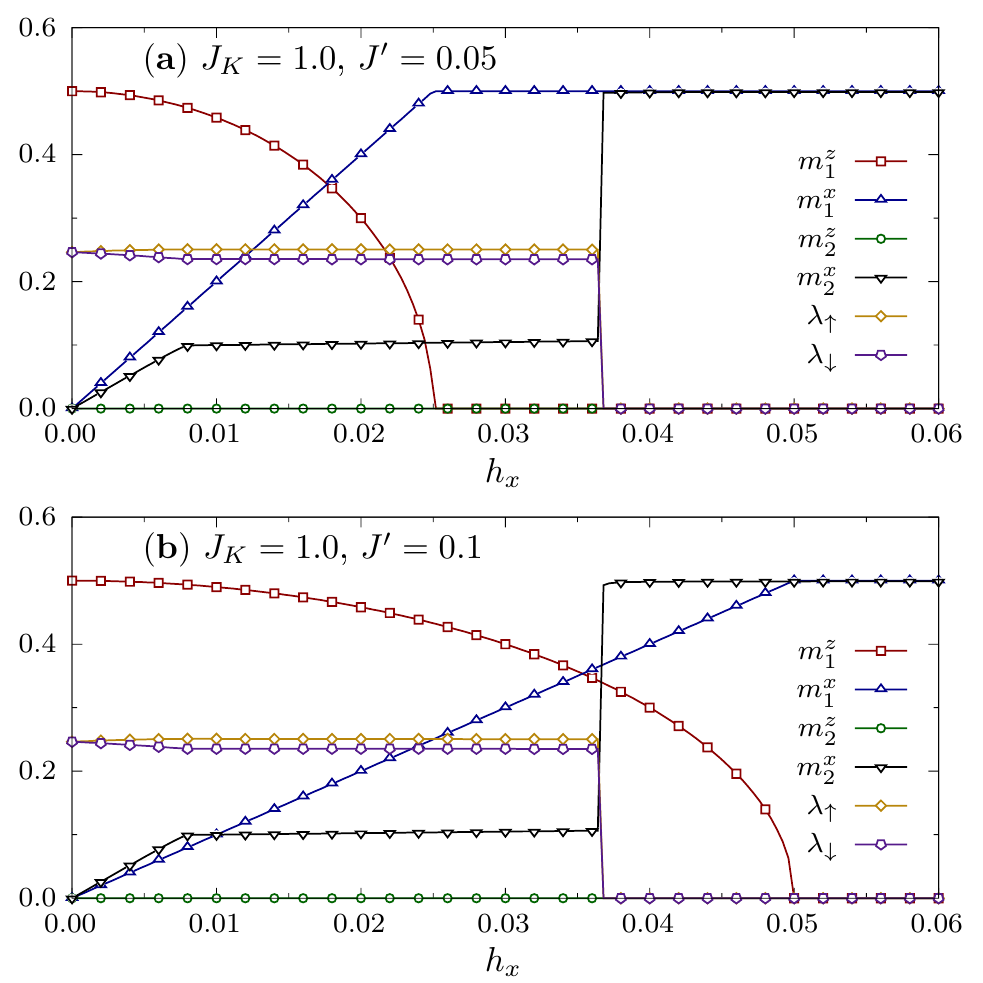}
\caption{\label{fig:jk1jh0jivaria} Variation of magnetization and Kondo coupling as a function of $h_x$ for different values of $J'$ with $J_K=1.0$ and $J_H=0$. {\bf (a)} $J'=0.05$. {\bf (b)} $J'=0.1$. The inclusion of Ising-like interaction gives values of $\vtr{m_{1}}$ different from zero for $h_x=0$, but it does not affect what happens in orbital 2. Increasing $J'$ increases the critical value where $m_{1z}$ goes to zero, and consequently, the region where $m_{1x}$ is different of its saturated value. }
    \end{figure}

Here, the effect of a magnetic field along the $x$ direction is analyzed by considering also the intersite Ising-like interaction $J'$ for the orbital 1. The Hund's coupling is not considered here, $J_H=0$. 
Therefore the orbital 1 is fully decoupled from the other electrons which, on their side, form a Kondo lattice system. The model for the orbital 2 and the conduction electrons is similar to the one discussed in section~\ref{Section:result-field-effect} and we thus choose here to fix $J_K=1.0$ which correspond to $h_{K}^{\star}\approx 0.036$. The evolution of the magnetizations of the two orbitals as well as the $f-c$ effective hybridization as a function of the magnetic field $h_x$ are depicted in figure~\ref{fig:jk1jh0jivaria}.  
We consider that the intersite magnetic exchange between electrons in orbital 1 is of the same order of magnitude as the critical field $h_{K}^{\star}$. The resultas for $J'=0.05$ in ~\ref{fig:jk1jh0jivaria}({\bf a}), and $J'=0.1$ in ~\ref{fig:jk1jh0jivaria}({\bf b}) are presented in figure~\ref{fig:jk1jh0jivaria}. 

The Kondo lattice formed by electrons in the orbital 2 is discussed in detail in section~\ref{Section:result-field-effect} with the  three regimes, linear, plateau, and saturation. 
We now focus on the analysis of orbital 1, which is saturated to its maximal value 
$\|\vtr{m_1}\|=1/2$, but not necessarily in the same direction as $m_{2}$. 
The Ising-like interaction $J'$ favors a magnetization along the $z$ direction, while the transverse magnetic field favors alignment along $x$. 
By exploiting the fact that orbital 1 is decoupled from other electrons we can solve exactly the mean-field equation for 
$\vtr{m_1}$ at zero temperature. Indeed, for $J_{H}=0$ and with a field oriented along $x$ axis, Eq.~\ref{eq:th1} has 
two possible solutions: either $m_1^{z}=0$ and $m_1^{x}=1/2$, which is realized at sufficiently large field. Or 
$m_1^{x}=h_x/J'$ and $m_1^{z}=\sqrt{\frac{1}{4}-\left(\frac{h_x}{J'}\right)^2}$ at fields $h_x$ lower than a critical 
value $h_M^{\star}=J'/2$. This linear increase of $m_1^{x}$ with $h_x$ is depicted in figure~\ref{fig:jk1jh0jivaria}{\bf (a)}, corresponding to a gradual rotation of the magnetization until its complete alignment along $x$ axis at $h_x\ge h_M^{\star}$. 
The model parameters used for the plots in figure~\ref{fig:jk1jh0jivaria}{\bf (a)} correspond to $h_M^{\star}=0.025$, which is slightly lower than the other critical field $h_K^{\star}\approx 0.036$ characterizing the vanishing of the Kondo $f-c$ hybridization and the full magnetization of orbital 2. However, by increasing the value of the Ising interaction $J'$, the alignment of $\vtr{m_1}$ along $x$ axis can also be obtained for higher critical field, inside the non-Kondo regime ($h_M^{\star}>h_K^{\star}$), as can be seen in figure~\ref{fig:jk1jh0jivaria}{\bf (b)} . 

\subsection{Effect of Hund's coupling}\label{subsec:hund}

We now extend our study by considering the effect of the transverse magnetic field in the presence of local Hund's coupling and intersite Ising interaction. In this case, orbital 1 and 2 interact by the Hund's coupling and orbital 2 is coupled with the conduction electrons by Kondo interaction. In the presence of magnetic field, the magnetization of orbital 2 will no longer be in the field direction because it is coupled to the magnetization of orbital 1. Moreover, the Weiss mean-field resulting from Hund's coupling will  add to the applied field and it is expected to weaken the effective Kondo hybridization 
, destroying it for a critical field smaller than ${h_K^{\star}}$ defined above.  

\begin{figure}[H]
  \centering
  \includegraphics[width=1\columnwidth]{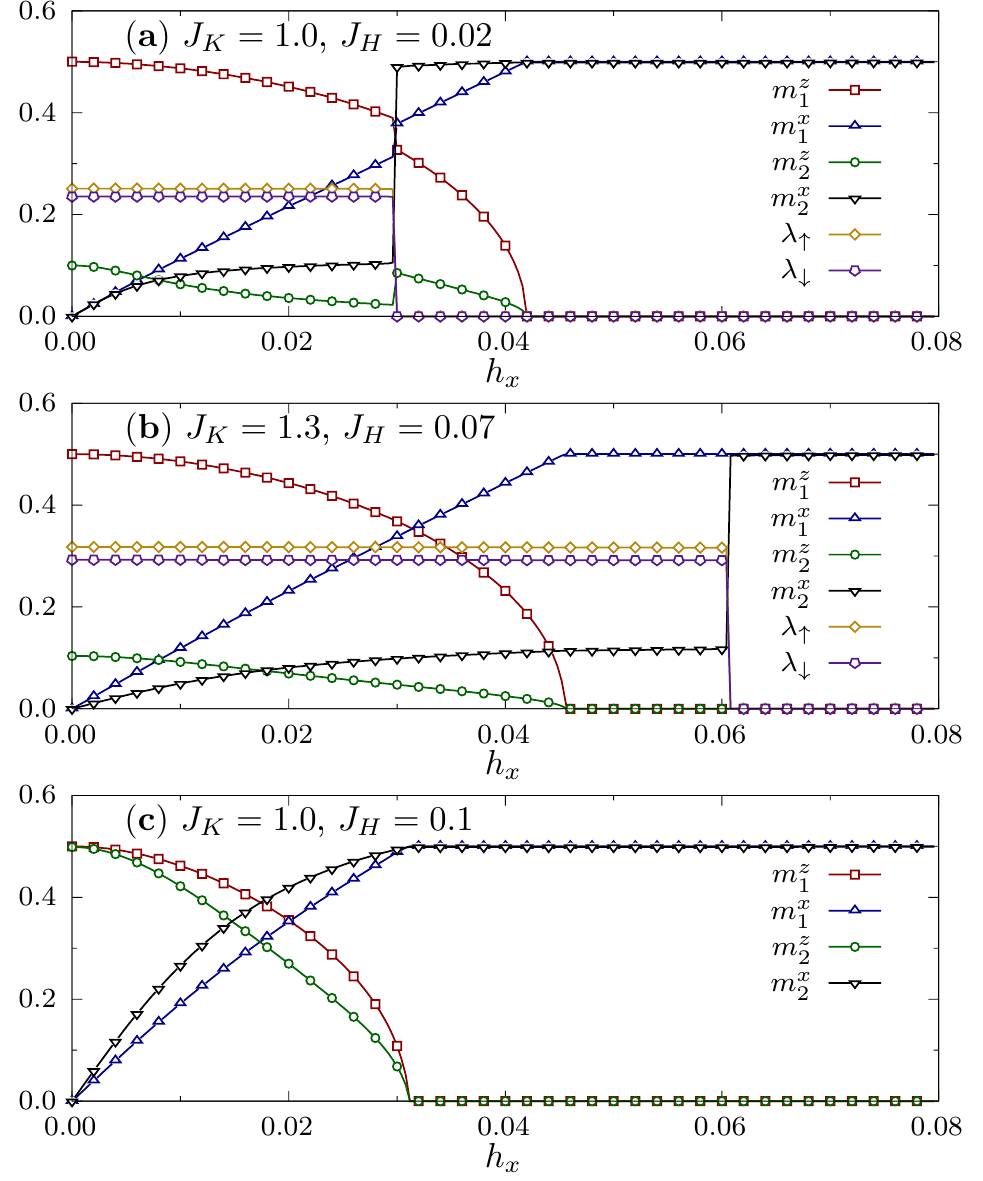}
  \caption{\label{fig:t0_jk1_jhvaria_ji01} Variation of magnetization and Kondo coupling as a function of $h_x$ for different values of $J_K$ and $J_H$ for fixed $J'=0.1$. $(\mathbf{a})$ $J_K=1.0$ and $J_H=0.02$ ; $h_K^{\star}<h_M^{\star}$ in this case. ($\mathbf{b}$) $J_K=1.3$ and $J_H=0.07$; $h_K^{\star}>h_M^{\star}$. ($\mathbf{c}$) $J_K=1.0$ and $J_H=0.1$. When $J_H$ is tuned, all parameters are linked. The Hund's coupling allows ta non-zero component of magnetization in $z$ direction for the orbital 2. This effect increases the global magnetization and can destroy the  Kondo coupling.}
\end{figure}

The evolutions of the various mean-field parameters as functions of the magnetic field $h_x$ are depicted in figure~\ref{fig:t0_jk1_jhvaria_ji01}. We consider two set of parameters corresponding to either  $h_M^{\star}>h_K^{\star}$ (Figure~\ref{fig:t0_jk1_jhvaria_ji01}$(\mathbf{a})$) or $h_K^{\star}>h_M^{\star}$ (Figure~\ref{fig:t0_jk1_jhvaria_ji01}$(\mathbf{b})$) described previously for $J_H$.

\begin{figure}[H]
\centering
\includegraphics[width=1\columnwidth]{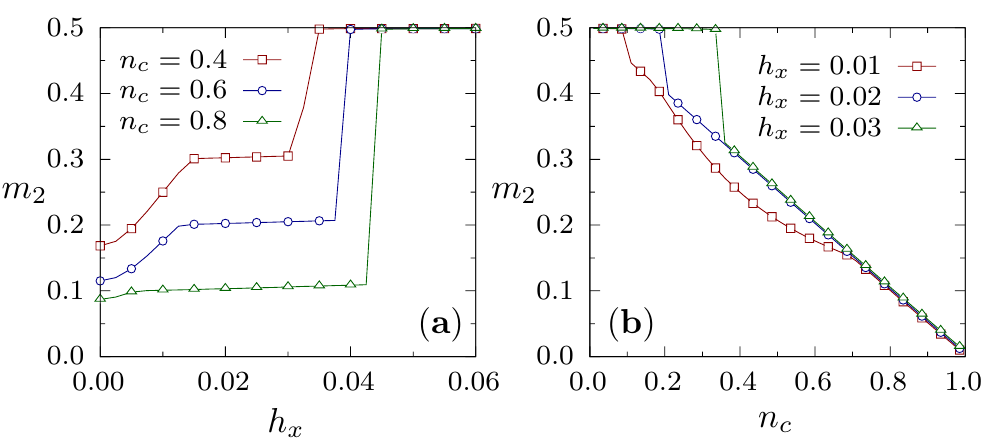}
\caption{\label{fig:m2xjk1j01jh001} $\mathbf{(a)}$ Variation of $m_2$ as function of $h_x$ for different values of conduction electrons and, $\mathbf{(b)}$ Variation of $m_2$ as function of the conduction electron number and different values of $h_x$ for $J_K=1.0$, $J'=0.1$, and $\jhu=0.02$.}
\end{figure}

The behaviors depicted in figure~\ref{fig:t0_jk1_jhvaria_ji01}{\bf (a)} for small $J_H=0.02$ is similar with the one depicted in figure~\ref{fig:jk1jh0jivaria}{\bf (b)} for 
$J_H=0$. For example, both plateau and saturated regimes are observed for the magnetization of orbital 2, but now for the modulus of $m_2$ (shown in Figure \ref{fig:m2xjk1j01jh001}~$\mathbf{(a)}$). Here, the small but non-zero $z$-component  comes from a combined effect of Hund's coupling with orbital 1 together with the Ising intersite interaction ; the variation with field is no longer linear behavior as it was for $J_H=0.0$. Also, some slight discontinuities are observed for the magnetizations $m_1^x$ and $m_1^z$ at the critical field $h_K^{\star}$, which marks the vanishing of the Kondo $f-c$ hybridization. 

The figure~\ref{fig:t0_jk1_jhvaria_ji01}~$\mathbf{(b)}$ obtained for $J_K=1.3$ and $J_H=0.07$ shows an increase of $h_K^{\star}$ since the Kondo 
coupling is increased. The discontinuities of $m_1^x$ and $m_1^z$ are not observed, and the $z$-component of $\vtr{m}_2$ is larger. Also, the stronger value of $J_K$ allows stabilization of the mixed state  with a magnetic moments of orbital 1 completely aligned to the magnetic field $h_x$. The figure~\ref{fig:t0_jk1_jhvaria_ji01}~($\mathbf{c}$) is obtained for $J_K=1.0$ and $J_H=0.1$. The resulting Weiss field induced by the fully polarized orbital 1 on the orbital 2 has stronger intensity than the previous cases reported before.The presence of Ising interaction together with Hund's coupling favors magnetizations aligned along $z$ direction for small $h_x$, while the alignment is along $x$ direction above a critical field $h_M^{\star}$. The rotations of the magnetizations induced by the field are continuous and gradual. 


The variation of the magnetization at orbital 2 as a function of the transverse magnetic field  and the number of conduction electron is depicted in figure \ref{fig:m2xjk1j01jh001}. For the three values of $n_c$ shown in figure \ref{fig:m2xjk1j01jh001}({\bf a}), we can see the presence of the plateau already discussed at $\sim(1-n_c)/2$. In figure \ref{fig:m2xjk1j01jh001}({\bf b}), the linear variation of $m_2$ is observed for some range of $n_c$, where the plateau is present.

\begin{figure}[H]
\centering
\includegraphics[width=1\columnwidth]{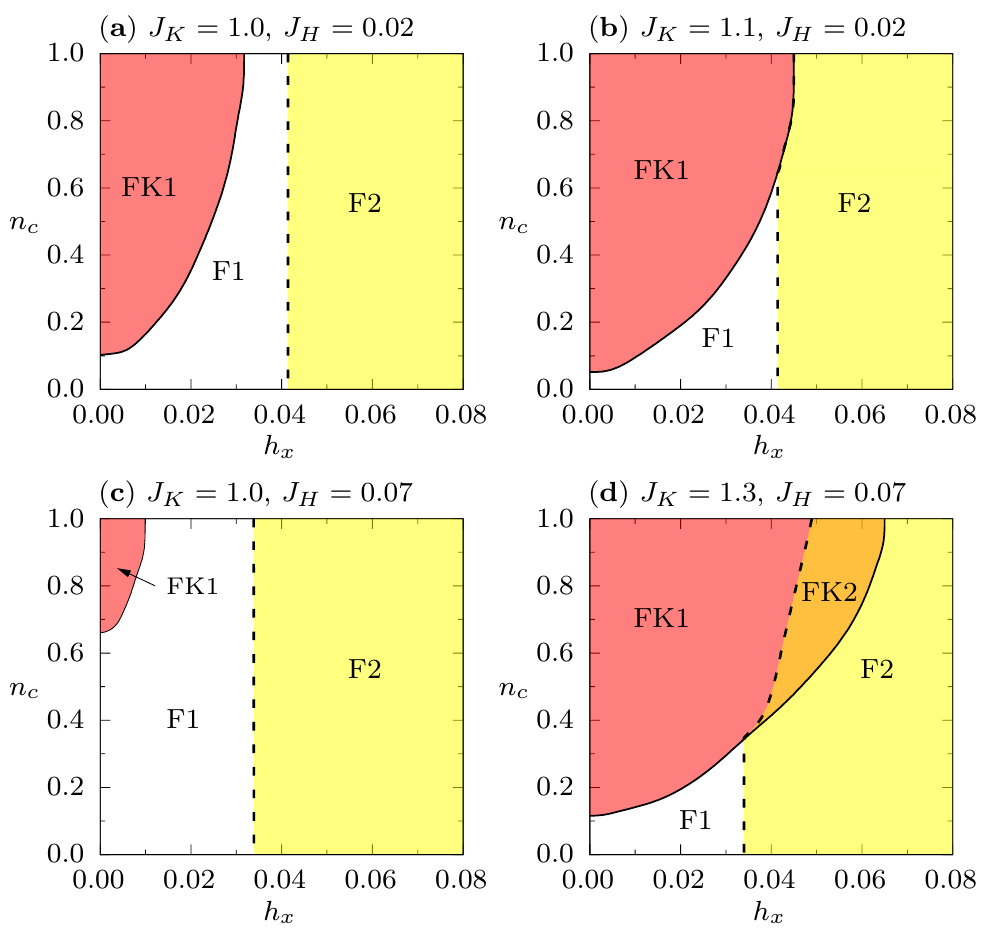}
\caption{\label{fig:phdia_nchx_j01jh001} Phase diagram as a function of the conduction electron occupation, $n_c$, and the transverse magnetic field, $h_x$, for $J'=0.1$, $T=0.0001$, and different values of $J_K$ and $J_H$. Depending on the interaction parameters  different phases are stabilized:  ferromagnetism and Kondo effect may compete or coexist. Strong values of the applied field always induce a fully polarized state, F2, with magnetization aligned along the field. 
At zero field, a ferromagnetic phase is always found due to the Ising interaction $J'$,  coexisting  with Kondo screening in the FK1 state  at sufficiently large $n_c$.  
{\bf (a)} and {\bf (c)} are obtained for relatively small $J_K=1.0$. Here, $h_K^{\star}<h_M^{\star}$ and the field-induced breakdown of Kondo effect, FK1-F1, is realized in a ferromagnetic phase where the magnetization is not fully aligned along the field. 
{\bf (b)} is obtained for a slightly larger value of Kondo interaction. Here, a direct transition FK1-F2 can be realized, reflecting $h_K^{\star}=h_M^{\star}$, and corresponding to a simultaneous breakdown of Kondo effect and allignement of the magnetizations along the field direction. 
{\bf (d)} is obtained for relatively stronger values of $J_K$ and $J_H$. Here, a cascade of transitions may be realized, FK1-FK2-F2. In the intermediate mixed Kondo phase FK2, the magnetization is aligned along the applied field.}
\end{figure}

The figure \ref{fig:phdia_nchx_j01jh001} shows the phase diagrams as a function of occupation number, $n_c$, and the transverse magnetic field, $h_x$, for fixed $J'=0.1$ and four combinations of $J_K$ and $J_H$. Following the description made in the subsection \ref{sec:j0jh0}, we can relate the different phases with the schematic pictures presented in figure \ref{fig:sche_phases}. In the white part of Figure \ref{fig:phdia_nchx_j01jh001}, the F1 phase is defined as a ferromagetic phase where the $z$ component of the magnetization is non-zero. In the yellow region, the F2 phase corresponds to a magnetization completely aligned with the applied field. It can be saturated for both orbitals, leading to the fully polarized state
. In the red region, FK1 indicates a mixed state: ferromagnetic (non fully polarized) order and Kondo hybridization coexist. Another mixted state is obtained, FK2, which appears in the orange region:  
here, the magnetization oriented along $x$ direction coexists with Kondo hybridization. 
The two diagrams represented by figures \ref{fig:phdia_nchx_j01jh001}$\mathbf{(a)}$ and $\mathbf{(c)}$, which mimic two different values of $J_H$, represent qualitatively similar situations: in both cases, FK1 can be present for sufficiently large value of electronic filling $n_c$ and  small field; this phase is limited by the critical field $h_K^{\star}$ above which the field always abruptly destroys the Kondo hybridization. Also, here, an increase of $J_H$  results in a  decrease of $h_K^{\star}$, which  vanishes at small $n_c$ where non Kondo state can be formed. In all cases, a fully polarized non-Kondo state F2 is realized for strong applied manetic field. 
We now focus onto the diagrams obtained for larger values of Kondo interaction. 
In \ref{fig:phdia_nchx_j01jh001}$\mathbf{(b)}$, an intermediate situation is observed with $J_K$  slightly stronger than the one in \ref{fig:phdia_nchx_j01jh001}$\mathbf{(a)}$. In this case, a direct transition from FK1 to F2 may be realized by applying a magnetic field if $n_c$ is sufficiently large. 
Such a FK1-F2 transition corresponds to $h_K^{\star}=h_M^{\star}$ and, as a consequence, all parameters are expected to vary abruptly, especially magnetization (value and orientation), electronic density of states, Fermi-surface. 
Figure \ref{fig:phdia_nchx_j01jh001}$\mathbf{(d)}$ depicts a situation obtained where both Kondo and Hund interactions are relatively stronger than in other figures. In this case, a cascade of transitions FK1-FK2-F2 can be induced by the magnetic field: as a first step, from FK1 to FK2, the magnetic field continuously turns the magnetization to $x$ direction, preserving continuously the non-zero Kondo hybridization just above $h_M^{\star}$. Then, in the intermediate Kondo phase FK2, the $x$ component of ${\bf m_2}$ reaches the plateau behaviour. The breakdown of Kondo effect can then be realized (FK2-F2 transition) at the critical field $h_K^{\star}> h_M^{\star}$.

\section{Discussion and conclusion}\label{Discussion}

The description presented in this paper is intended to shed light on the discussion of the metamagnetic transition in the URhGe compound. With this proposal in mind, we studied the effect of the transverse magnetic field in a model with two $f$-electron orbitals that interact with onsite and intersite exchange, and are in contact with a background of conduction electrons. Here, the interplay between Kondo effect and ferromagnetic order has a  strong influence on the properties of the system. 
We focused on the tunability of the ground state properties by applying a transverse external magnetic field. The two main possible effects that might be induced by applying a transverse magnetic field in this system are: breaking of the Kondo effect, and rotating the magnetization axis along the direction of the field. 
We obtained and characterized different scenarios for the predicted field-induced transitions. When the critical field $h_K^{\star}$ (field necessary to destroy Kondo effect) is smaller than $h_M^{\star}$ (field necessary to rotate the spins along $x$ axis) (see, e.g., Figure \ref{fig:phdia_nchx_j01jh001}{\bf (a)}), the Kondo effect is destroyed before the field completes the rotation of the magnetization along $x$.
In this case, we predict an abrupt transition from phase FK1 to phase F1 (at $h_K^{\star}$), followed by a continuous rotation of the magnetization as a function of the transverse field (ended at $h_M^{\star}$) up to the fully polarized F2 phase. On the other hand, when  $h_K^{\star}$ is bigger than  $h_M^{\star}$ (see, e.g., Figure \ref{fig:phdia_nchx_j01jh001}{\bf (d)}), the complete rotation of the spins occurs inside the Kondo phase ($m_{2z}$ is zero but $m_{2x}$ is not saturated) from phase FK1 to phase FK2 and the metamagnetic transition indicating the breakdown of Kondo effect is obtained only when the magnetization is already along the $x$ direction, from phase FK2 to F2. For an intermediate set of parameters (see, e.g., Figure \ref{fig:phdia_nchx_j01jh001}{\bf (b)}), the critical fields 
can coincide, $h_K^{\star}=h_M^{\star}$, 
 and the complete rotation of the magnetization is expected to occur at the same field as  the Kondo breakdown metamagnetic transition, from phase FK1 to phase F2.

These different situations, resulting from different values of the dimensionless ratio $h_K^{\star}/h_M^{\star}$, correspond to different scenarii, which are associated with different experimental signatures: focusing on the magnetic signatures, Kondo breakdown at $h_K^{\star}$ is expected to be revealed by a metamagnetic transition, while $h_M^{\star}$ marks the full rotation of the magnetization along the direction of the applied field. Analyzing magnetization curves as a function of transverse field should thus provide a clear way to discriminate between the different scenarii. Other signatures of Kondo effect, like the expected increase of the effective mass, also need to be analyzed coherently with the magnetization. 

Regarding the discussion of the experimental results in URhGe, we identify our theoretical parameter $h_M^{\star}$ with $H_R$ (spin-reorientation field). This is consistent with the phenomenological analysis given in~\cite{Mineev2021}, using Landau free energy expansion. 
Experimentally, an increase of the effective mass around $H_R$ 
is revealed in different contexts: it is reported by specific heat~\cite{Hardy2011} that $\gamma$ has a peak, and the constant $A$ of the $T^2$ relation on the resistivity, has similar behavior~\cite{Gourgout2016}.
Assuming the Kadowaki-Woods relation $\sqrt{A}\sim m^{*}$ and considering that the effective mass has two main contributions, a band mass $m_B$ and a magnetic contribution $m^{**}$,  \(m^*=m_B+m^{**}\), it was proposed ~\cite{Hardy2011} that, when the effective mass increases, it is the magnetic part due to the magnetic fluctuations that is responsible of this increase. The band mass contribution is considered as field independent and does not change across the transition~\cite{Miyake2008}.
Also, the Fermi surface presents a variation of around 7\% detected by quantum oscillations (Shubnikov-Das Haas experiment) at $H_R$ and it is reported that the effective mass decreases when crossing  $H_R$ and stays constant for higher fields~\cite{Yelland2011}. Measurements of Hall effect also indicates a Fermi surface reconstruction~\cite{Aoki2014}, together with the results of thermoeletric power that changes sign around $H_R$~\cite{Gourgout2016}. ARPES shows itinerant behavior for the $5f$ electrons in the 
ferromagnetic state~\cite{Fujimori2014}, but there are no result above to the metamagnetic transition. 

 On the basis of these experimental results, we propose two possible descriptions : first, a scenario with $h_M^{\star}<h_K^{\star}$ as depicted in Figure \ref{fig:phdia_nchx_j01jh001}{\bf (d)}, can be realized if experimentaly 
 Kondo effect is established above and below $H_R$. However, in this configuration, we would not observe any abrupt variation in the magnetization at that field, and the metamagnetic transition might occur only for higher values of the transverse field when $h_x=h_K^{\star}$. The second scenario 
 corresponds to $h_K^{\star}=h_M^{\star}$ (see for example \ref{fig:phdia_nchx_j01jh001}{\bf (b)}), where the rotation is accompanied with a sudden variation of the magnetization along the applied transversal field. This could be the case if the effective mass has an abrupt change across the metamagnetic transition.

The  Das Haas-van Alphen experiment on URhGe could help in  understanding the nature of the metamagnetic transition discussed here, analyzing how the effective band mass changes across $H_R$. Finally, we did not take into account any possible effect on the metamagnetic transition due to the reentrant superconducting phase, but we strongly believe that the interaction of the itinerant and localized electrons is important for the description of the magnetic and superconducting effects in the uranium compounds. 

\section*{ACKNOWLEDGMENTS}

C.T.~acknowledges partial  support provided by the Brazilian-France Agreement CAPES-COFECUB (No.~88881.192345/2018-01). We would like to thank the COTEC (CBPF) for making available their facilities for the numerical calculations on the \textit{Cluster HPC}.

\bibliography{magkondo2fc.bib}{}
\bibliographystyle{elsarticle-num}

\end{document}